# Thermo-Optical Tuning of Whispering Gallery Modes in Er:Yb Doped Glass Microspheres to Arbitrary Probe Wavelengths

Amy Watkins,[1,2] Jonathan Ward,[1,3] and Síle Nic Chormaic[1,2,4,5]

[1]*Physics Department, University College Cork, Cork, Ireland*

[2]*Photonics Centre, Tyndall National Institute, Lee Maltings, Cork, Ireland*

[3]*Nano-Optik, Institut für Physik, Humboldt Universität zu Berlin, 10117 Berlin, Germany*

[4]*School of Physics, Westville, University of Kwa-Zulu Natal, Durban 4041, South Africa*

[5]*Light-Matter Interactions Unit, Okinawa Institute of Science and Technology Graduate University, Okinawa 904-0411, Japan*

We present experimental results on an all-optical, thermally-assisted technique for broad range tuning of microsphere cavity resonance modes to arbitrary probe wavelengths. An Er:Yb co-doped phosphate glass (Schott IOG-2) microsphere is pumped at 978 nm via the supporting stem and the heat generated by absorption of the pump light expands the cavity and changes the refractive index. This is a robust tuning method that decouples the pump from the probe and allows fine tuning of the microsphere's whispering gallery modes. Pump/probe experiments were performed to demonstrate thermo-optical tuning to specific probe wavelengths, including the $^5S_{1/2}$ F = 3 to $^5P_{3/2}$ F' = 4 laser cooling transition of $^{85}$Rb. This is of particular interest for cavity QED-type experiments, while the broad tuning range achievable is useful for integrated photonic devices, including sensors and modulators.

Email address: amy.watkins@tyndall.ie

# 1. Introduction

Whispering gallery mode (WGM) resonators have been used comprehensively in a range of areas from the fundamental, in cavity-QED experiments[1-3] and cavity optomechanics,[4,5] to applied physics in bio- and chemical sensing,[6-11] or laser engineering.[1,12-16] One key requirement for either a sensing or a cavity-QED application is the ability to tune the microcavity resonances to match the energy transition of the atomic or molecular species under investigation. Many tuning techniques for the WGMs in spherical microcavities have been proposed, including chemical etching,[17] heating the cavity via an external heater[18] and deforming the cavity under strain using stretching or squeezing.[19-21] However, there are disadvantages associated with each of these techniques. Chemical etching is an irreversible process and the WGM resonances cannot be scanned around a fixed frequency. The stretching/squeezing method has a wide tuning range but, due to the relatively complex mechanical structures involved, it is difficult to realize experimentally and fine tuning may also be a problem. In contrast, tuning using an external heater has only been used for fine tuning experiments. Electro-optical tuning of the resonant frequencies of microtoroidal cavities at 1550 nm has been demonstrated,[22] with tunings as large as 300 GHz being achieved. More recently, electrically tunable liquid crystal optical microresonators were tuned over 20 nm by applying an external voltage.[23] Furthermore, small thermo-optically induced resonance shifts of ~100 GHz have previously been observed in erbium-doped fluoride glass (ZBLALiP) microspheres pumped at 1480 nm.[24]

In this paper, we report on an all-fiber, thermo-optically assisted method of tuning the WGM resonances of Er:Yb co-doped glass microspheres to specific probe wavelengths. Initially, the WGMs are excited through coupling with probe light in the evanescent field of an adiabatically tapered optical fiber[25] while tuning of the modes is achieved by pumping the resonators at 978 nm via a supporting structure i.e., a stem, as shown in Fig. 1. Absorption of this pump light leads to internal heating of the resonator, with the heat concentrated in the WGM volume. Accordingly, the transmitted power in the tapered fiber can be kept to a minimum which is favorable for a wide range of applications e.g., under vacuum to avoid any heat damage. By pumping up through the stem, we also avoid coincidental pump/cavity resonances which can cause a sudden increase in cavity temperature and a large redshift of the cavity modes. However, this would not be the case if the pump were propagating in the tapered fiber as this would result in non-linear tuning of the WGMs.[26] By simply changing the power of the 978 nm pump sent through the stem, linear tuning ranges up to 488 GHz have been reported.[26] Here, we demonstrate resonant tuning of the cavity to a 1300 nm probe laser and, additionally, to the $^5S_{1/2}$ F = 3 to $^5P_{3/2}$ F' = 4 cooling transition of $^{85}$Rb for extended periods on the order of tens of seconds.

## 2. Experimental Setup

The microspheres are made from Schott IOG-2 laser glass doped with 2 wt% $Er_2O_3$ ($1.7 \times 10^{20}$ ions/cm$^3$) and co-doped with 3 wt% $Yb_2O_3$ ($2.5 \times 10^{20}$ ions/cm$^3$).[27] The operation of Er-doped microsphere lasers emitting in the C- and L-bands has been discussed extensively in the literature.[28-30] A schematic of the experimental setup is shown in Fig. 1. A microsphere is glued (using UV epoxy) to a 1 mm long glass stem made by heating and pulling standard optical fiber using a $CO_2$ laser. The diameter of the tip of the stem is 20 μm. The other end of the fiber is spliced to a 978 nm FBG diode laser, which is used to send pump light up through the stem. The 978 nm pump power is adjusted using an electronically-controlled variable attenuator or by changing the laser diode current using a function generator.

The 978 nm pump light entering the sphere via the stem does not form strong WGMs and only a small fraction of the launched pump light couples into the tapered optical fiber containing the probe. This can be filtered out using a fiber-coupled wavelength division multiplexer (WDM). In our setup two different lasers (details in §3) are used as probes depending on the experiment. As stated earlier, efficient coupling into the microsphere is achieved via adiabatically tapered optical fibers.[25] The tapered fibers are fabricated using a direct-heating technique with an oxy-butane flame and they are in contact with the microsphere in all of the experiments. This improves the stability of the system at the cost of a slightly reduced quality (Q) factor due to an increase in loss mechanisms. We have measured minimum Q factor values of ~$10^5$. The transmission through the tapered fiber is recorded on a digital oscilloscope using an InGaAs photodiode.

## 3. Thermo-Optical Tuning of the Microcavity

In earlier work,[31] we showed that the thermal limit of Er:Yb IOG-2 microspheres could be reached by pumping the sphere at 978 nm via a tapered fiber. Here, the pump light through the tapered fiber is replaced by a probe laser (linewidth <200 kHz) centered at 1300 nm. A 100 μm IOG-2 sphere is placed in contact with the taper and the 978 nm pump power in the stem is modulated back and forth at 0.5 Hz from zero to full power (~300 mW) using a linear ramp, while the 1300 nm transmitted power is recorded on the oscilloscope. Figure 2 shows a typical scan of the cavity as the 978 nm pump power in the stem is increased. The dips in the transmitted probe power correspond to cavity/probe resonances. The oscilloscope can be triggered on the rising edge of the modulating voltage ramp so that the same cavity/probe resonances can be continuously monitored. Next, the probe wavelength was increased from 1316.10 nm to 1316.15 nm and the measurement was repeated. The resulting time shift between the two recorded scans was used to convert to a frequency variation. The average tuning range[26] of the sphere is ~300 GHz at 1316 nm, with estimated Q factors of $10^5$. Q factors as high as $10^7$ have been measured using this probe. The thermal effect on the resonance wavelength shift, $\delta\lambda$, can be estimated from[24, 26, 28] such that

$$\delta\lambda = \lambda\left(\frac{1}{n}\frac{\delta n}{\delta T} + \frac{1}{d}\frac{\delta d}{\delta T}\right)\delta T, \qquad (1)$$

where $\delta T$ is the increase in microsphere temperature due to the pump laser, $\delta d$ is the longitudinal expansion of the microcavity and $\delta n$ is the change in refractive index of the material. Also, $\lambda$ is the resonance probe wavelength and $d = 2\pi R$, where $R$ is the radius of the sphere. The thermal expansion of the IOG-2 glass is $145 \times 10^{-7}$ K$^{-1}$ [32] while the refractive index change is estimated to be $-10 \times 10^{-7}$ K$^{-1}$.[28] Thus, in Eq. 1, the expansion parameter is dominant since it is about 10 times larger than that of the refractive index, which can be considered negligible on the resonance wavelength shift of the cavity. The mode shift of 0.05 nm in Fig. 2 corresponds to a total shift rate of 0.019 nm/K. This is comparable to the value of 0.022 nm/K evaluated using measured thermal shift of the lasing modes.[26]

The resonances highlighted in Fig. 2 are isolated by reducing the amplitude and DC offset of the voltage ramp driving the 978 nm laser diode current until the required number of cavity/probe resonances is observed, as shown in Fig. 3. The same resonances appear at different positions and with different shapes on the rising and falling edge of the pump ramp due to bistability in the cavity.[33-35] This is caused by positive and negative thermal feedback as a function of resonance detuning[34] and the amount of launched probe power can affect the shape and apparent linewidth of the resonances.[35] For low probe powers (less than −3 dBm) the cavity/probe resonance appears broad and relatively symmetric because the thermal feedback is small and the resonances have the same shape and position on both sides of the ramp. In contrast, for higher probe powers, the cavity/probe resonance produces enough heat to accelerate the cavity scan rate, thereby making the resonance appear narrower on the rising edge of the ramp and broader on the falling edge of the ramp, see Fig. 3. At probe powers greater than 9 dBm the cavity/probe resonance does not appear to narrow any further. The ramp amplitude can be reduced further and the DC offset adjusted until a single cavity/probe resonance is observed, as shown in Fig. 3(b). The stability of the mode is investigated by observing its position on the ramp over time. For repeated scans, the position of the dip only varied by less than one or two linewidths. This is quite good considering that the experiments were performed inside a basic isolating enclosure. The mode position on the ramp shifts if the microsphere/taper system is perturbed or the wavelength of the probe is changed.

## 4. Tuning into Resonance

So far, we have shown that the cavity can be scanned across single cavity/probe resonances. However, in most practical applications, it is necessary to tune the cavity to the resonance and for it to remain there. Tuning the cavity into resonance on the rising edge of the ramp (increasing pump power) when the probe power is high is very difficult due to unstable thermal feedback.[34] However, tuning close to the resonant peak is possible on the falling edge of the ramp because the thermal feedback creates a self-stabilizing cavity.[34] Alternatively, the probe power can be reduced so that thermal feedback is negligible. In either of these regimes, the cavity can be tuned to the bottom of the resonance dip by increasing or decreasing the 978 nm pump power in the stem.

Figure 4 shows the transmitted 1300 nm probe power when the pump power is changed near a cavity/probe resonance. The probe power was kept low to avoid thermal feedback in the mode. Having isolated a resonance of interest, the 978 nm pump power ramp was stopped just before the resonant dip. The pump power was then slowly increased until the resonance was observed. Next, the pump power was slowly decreased until the transmitted probe power dropped to a minimum, thereby indicating that the cavity is on resonance with the 1300 nm probe. To scan the cavity in and out of the resonance the pump power is changed by approximately ± 1.5 mW. The resonance condition is very sensitive to changes in the coupling condition and the external environment. Generally, the cavity will stay on resonance for a few tens of seconds; this is quite long considering the experiment is performed in a noisy laboratory inside a basic enclosure. A change in the resonance condition is observed as an increase in the transmitted power (see the noise region in Fig. 4). The noise is applied through vibrations on the optical table which in turn disturb the cavity-taper coupling. When the vibrations cease, the cavity returns to resonance with the probe. By varying the pump power it is possible to tune back out of the resonance.

Next, the 1300 nm probe laser is replaced by a single mode 780 nm laser (Toptica DL100) with a linewidth of ~1 MHz in an ECDL configuration. For this experiment, the tapered fiber was replaced with one that is designed to be single mode for 780 nm propagating light. This 780 nm probe laser is locked to the strong dipole (cooling) transition $^5S_{1/2}$ F = 3 to $^5P_{3/2}$ F' = 4 of the $^{85}$Rb isotope via a saturated absorption signal and a lock-in amplifier. As in the experiment with the 1300 nm probe, the 978 nm pump is modulated so that it scans back and forth across a number of cavity/probe resonances for a sphere with a diameter of 56 μm, as shown in Fig. 5(a). Then the unmodulated pump power is slowly adjusted so that the sphere is tuned to resonance with the 780 nm probe (see Fig. 5(b)). Here, the noise is absent as no external vibrations were applied to the system. The cavity remained on resonance with the 780 nm probe for an extended period of time on the order of 100 seconds. For comparison, in a typical laser-cooling experiment, the 1/*e* loading time for magneto-optical traps of different atomic species is less than or on the order of 2 seconds[36-39] and therefore 100 seconds would provide ample time for atom-cavity type experiments.

## 5. Conclusion

We have demonstrated the ability to tune laser glass microspherical cavities to arbitrary probe wavelengths by sending 978 nm pump light through the supporting stem of a microsphere. Thermo-optical tuning of ~117 GHz was achieved at a probe wavelength of 1316 nm and Q factors on the order of $10^5$ were observed even though the probe wavelengths are near the material absorption lines. The total shift rate observed is comparable to the rate estimated by measuring the shift of the microspheres lasing modes.[26] The pump light sent into the microsphere via the stem does not form coherent WGMs and is weakly coupled to the tapered fiber, as a result, any pump/cavity resonances which can cause sudden and sharp increases in temperature, were avoided. Thermal feedback in the probe/cavity resonances can be controlled by the probe power.

As an application, we have tuned the cavity to the $^{85}$Rb cooling transition. By varying the pump power through the supporting stem the cavity was tuned into resonance with the cooling transition for timescales on the order of 100 seconds. This method of tuning via the supporting stem of the sphere is a promising technique for vacuum-compatible experiments as the tapered fiber alone cannot propagate more than a few microwatts under vacuum before irreversible damage occurs in the form of melting. Recently, a non-destructive temperature measurement on $^{85}$Rb atoms in a magneto-optical trap (MOT) was reported using a passive (i.e., no probe beam) optical nanofiber[40] and sub-Doppler temperatures were observed, despite the presence of the nanofiber. We propose that the addition of a microsphere into a cold atom system would also have a negligible effect on the properties of a cold atomic cloud due to the small dimensions of the microsphere compared to the cloud. This is the focus of future work with the intention of investigating atom-cavity interactions.

Throughout these experiments, the linewidth of the microcavity (~GHz) is several orders of magnitude larger than the probe laser (<200 kHz or ~1 MHz depending on laser used). It would be more challenging to thermo-optically tune a narrow linewidth cavity (~kHz) into resonance with a laser for a substantial period. However, even with a $Q$ factor of $10^7$, which corresponds to the upper limit for a doped microcavity due to ion induced defects,[41] the linewidth of the doped cavity will still be a few orders larger than the laser. What we demonstrate here is a robust tuning method that decouples the pump from the probe. The pump can be low quality but high power and the probe can narrow linewidth while reasonable microcavity quality factors are still obtainable.

In the work reported here, Er:Yb co-doped microsphere cavities were used. However, different dopants and doping concentrations could be used to give optimal quality factors and tuning ranges. It may also be possible to thermally tune undoped, passive optical microcavities by placing them, or growing them, on top of a doped microcavity or thermo optical pad. The doped cavity or pad could be optically pumped and would transfer heat to the undoped cavity. This also forms the focus of future work.

## Acknowledgements

This work was partially supported by Science Foundation Ireland under Grant No. 07/RFP/PHYF518s2. AW acknowledges support from IRCSET through the Embark Initiative and OIST Graduate University. JW acknowledges support from IRCSET through the INSPIRE program.

# Figure Captions

**Fig. 1.** (Color online) Schematic of the experimental setup. Inset: Image of the microsphere with visible 978 nm light pumped via the stem. WDM: wavelength division multiplexer, VA: electronic variable attenuator, SMF: single-mode fiber, DSO: digital storage oscilloscope.

**Fig. 2.** (Color online) Scan showing the shift of the cavity/probe resonances over time as the pump power in the stem is increased from 0 to 300 mW. Changing the probe wavelength by 0.05 nm enables the conversion from the time axis to a frequency axis. The negative time is the storage time of the DSO trigger. The highlighted section represents resonances that are studied in further detail.

**Fig. 3.** (Color online) (a) The multiple resonances highlighted in the box in Fig. 2. (b) A single resonance from this group is isolated by changing the pump modulation frequency, offset and amplitude. The maximum pump power used is 9 dBm. To restore the symmetry of the resonance shapes and to reduce the bistability a probe power below $-3$ dBm is used.

**Fig. 4**. (Color online) Tuning the cavity to resonance with the 1300 nm probe laser. The noise was created by tapping lightly on the optical table. The dashed line shows the direction of the 978 nm pump power.

**Fig. 5**. (Color online) (a): Transmitted and scattered 780 nm probe light from a 56 μm sphere as 978 nm pump power is modulated near a cavity/probe resonance. (b) Tuning a 56 μm sphere to resonance with a 780 nm probe by adjusting the pump power in the sphere stem. The cavity stays on resonance for more than 100 seconds.

FIGURE 1

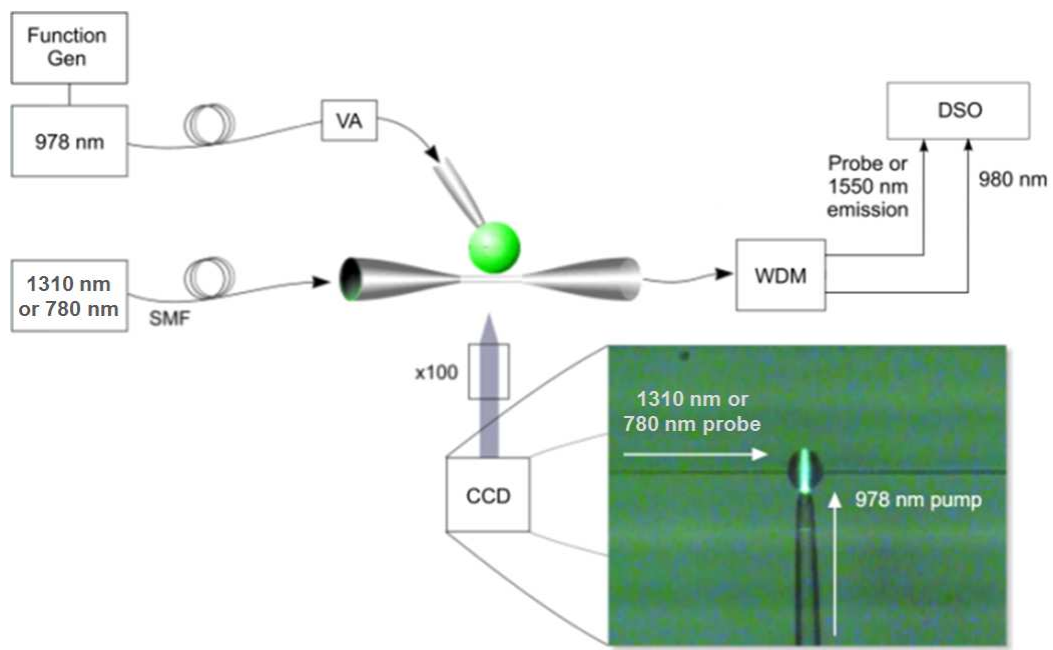

FIGURE 2

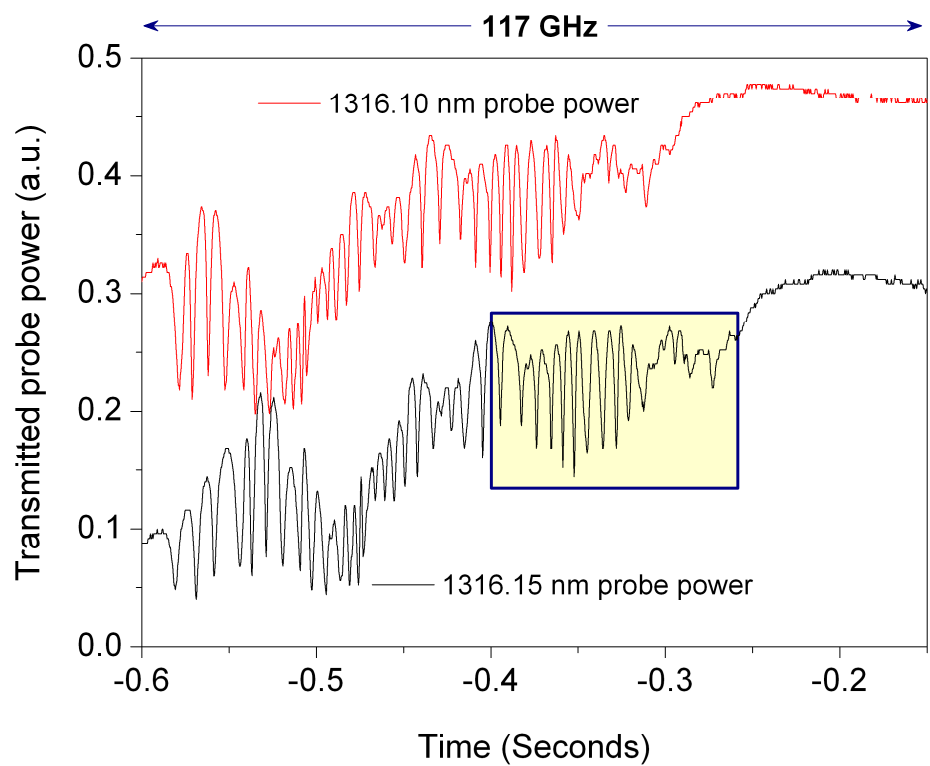

FIGURE 3

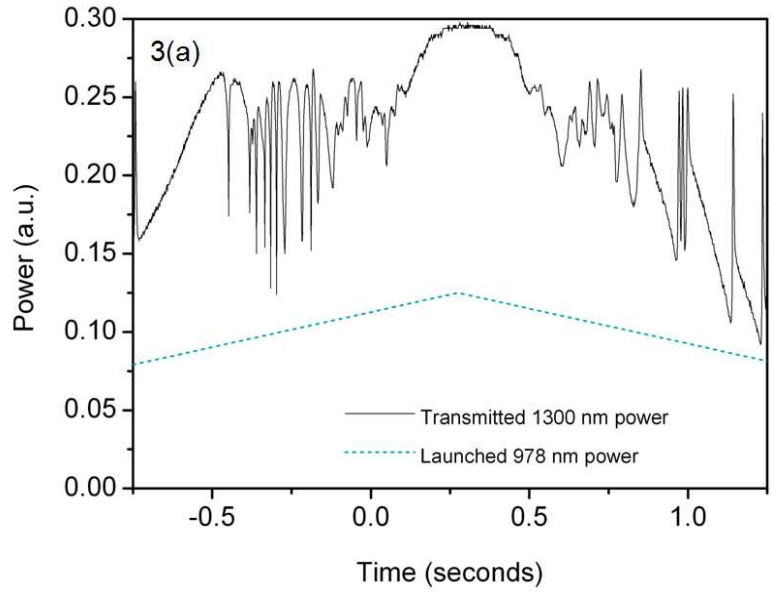

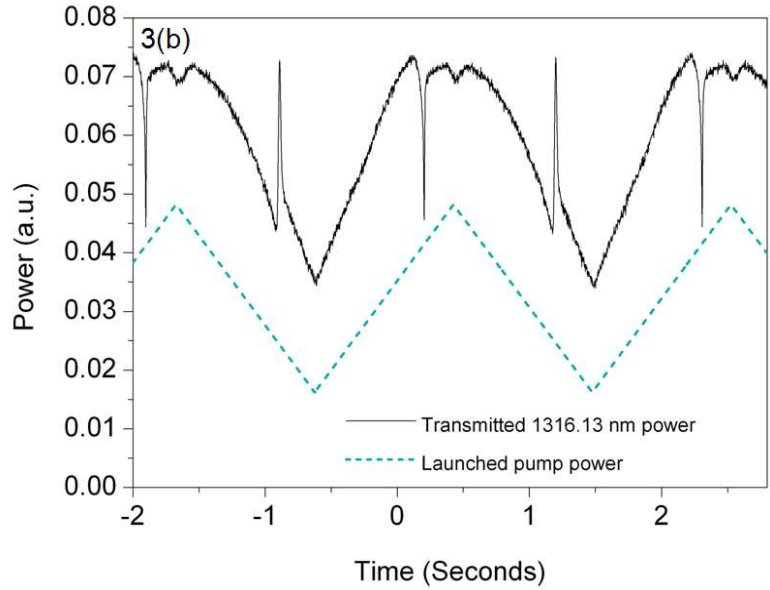

FIGURE 4

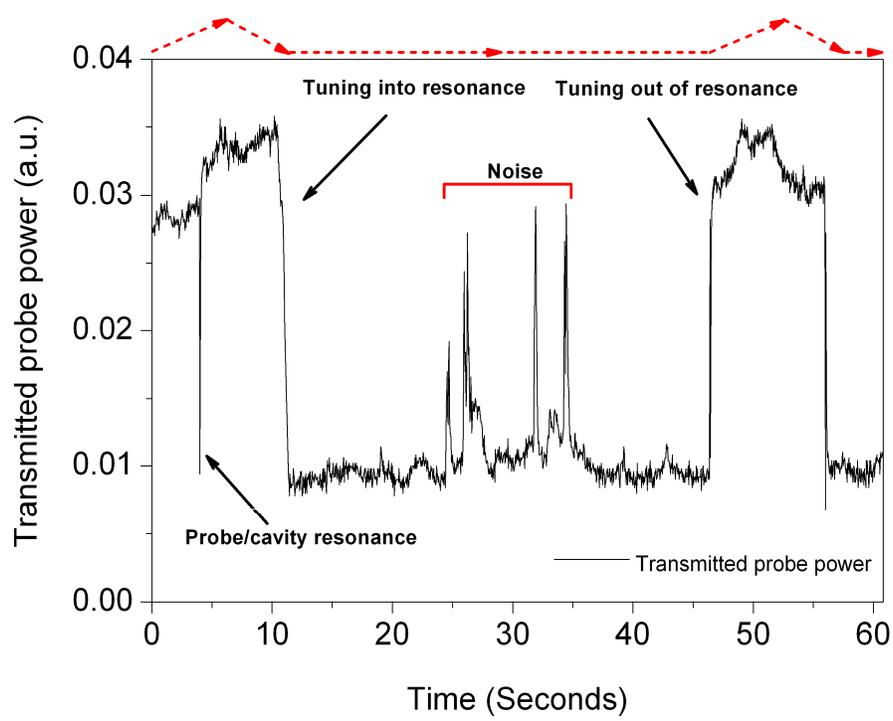

FIGURE 5

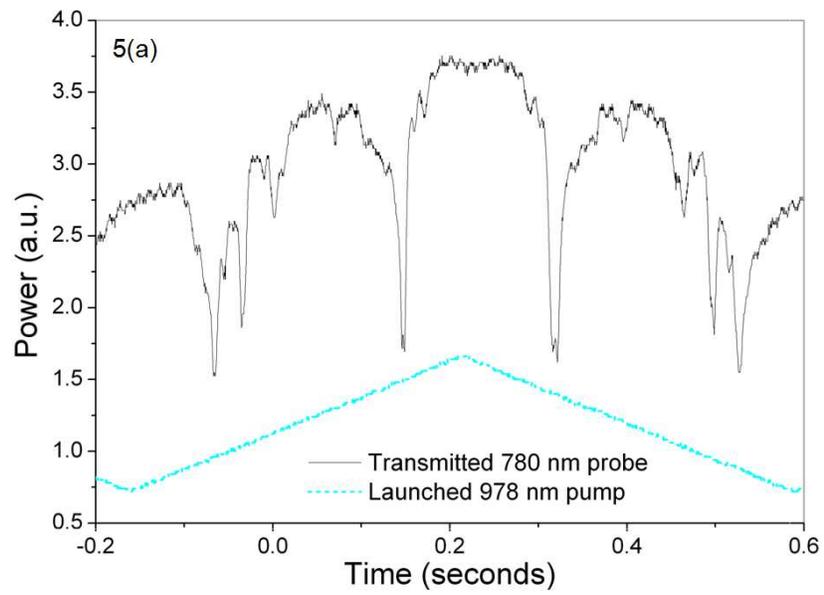

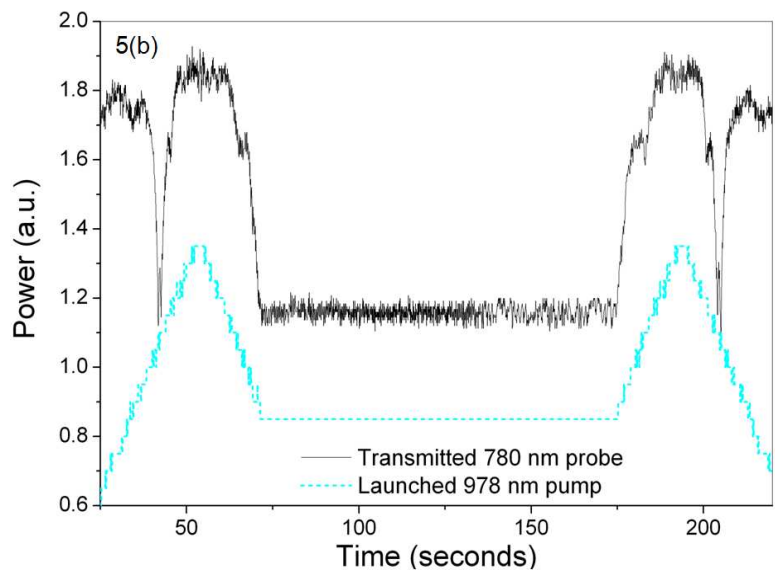